\documentclass[aps,prl,twocolumn,showpacs]{revtex4}
\usepackage{epsfig}
\usepackage{graphicx}

\begin{document}

\DeclareGraphicsExtensions{.eps,.EPS}

\title{Thermodynamics of a Bose Einstein condensate with free magnetization}
\author{B. Pasquiou, E. Mar\'echal, L. Vernac, O. Gorceix and B. Laburthe-Tolra}
\affiliation{Laboratoire de Physique des Lasers, UMR 7538 CNRS,
Universit\'e Paris Nord, 99 Avenue J.-B. Cl\'ement, 93430
Villetaneuse, France}

\begin{abstract}
We study thermodynamic properties of a gas of spin 3 $^{52}$ Cr atoms across Bose Einstein condensation. Magnetization is free, due to dipole-dipole interactions (DDIs). We show that the critical temperature for condensation is lowered at extremely low magnetic fields, when the spin degree of freedom is thermally activated. The depolarized gas condenses in only one spin component, unless the magnetic field is set below a critical value, below which a non ferromagnetic phase is favored. Finally we present a spin thermometry efficient even below the degeneracy temperature.

\end{abstract}

\date{\today}
\maketitle

A multi-component Bose Einstein condensate (BEC) (i.e. a spinor BEC) reveals ground and excited-states properties very different from scalar BECs \cite{Ho,Machida0}. Optical traps allow the study of such systems as all Zeeman states are trapped with almost the same trapping potential. New quantum phases arise due to an interplay between spin dependent contact interactions, linear and quadratic Zeeman effects \cite{Ueda review}. While these phases have been studied in a pioneering set of experiments \cite{spinor}, the properties at non zero temperature remain mostly unexplored (see however \cite{StamperKurn},\cite{Sengstock}). In addition, DDIs, up to now neglected in this context may lead to new quantum phases and non-trivial spin textures \cite{Ueda review}.

In this paper, we study the thermodynamic properties of a multi-component gas, made of spin $S$=3 chromium (Cr) atoms, at low temperatures. Our system strongly differs from alkali gases due to strong DDIs. One important consequence in the context of spinor gases is that  magnetization is free \cite{PasquiouPRA}. We operate at extremely low magnetic fields ($g_J\mu_B B \approx k_BT$, with $\mu_B$ the Bohr magneton, $k_B$ the Boltzmann constant, $T$ the temperature and $g_J=2$ for Cr) which results in non-zero population in excited Zeeman states at thermal equilibrium. We therefore investigate the phase diagram of spinor bosons with free magnetization.

\begin{figure}[h]
\centering
\includegraphics[width=3.2in]{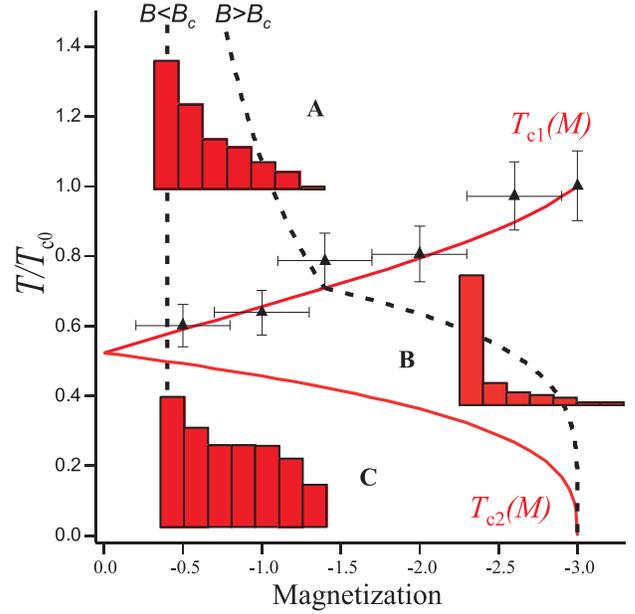}
\caption{\setlength{\baselineskip}{6pt} {\protect\scriptsize (Color online) Phase diagram of a spin 3 BEC. The solid lines delimitate the 3 phases predicted for a non interacting gas of bosons. A phase: thermal gas in each Zeeman component; B phase: BEC only in $m_S=-3$; C phase: BEC in all Zeeman components. The histograms represent typical experimental population distributions. Black triangles are experimental $T_{c1}$ measurements (see text). The dashed lines apply to regimes below $B_c$ (left) or above $B_c$ (right). }} \label{PhaseDiag}
\end{figure}

Our main results are summarized in Figure \ref{PhaseDiag}. At large temperatures,  we observe a normal (A) phase. For large enough magnetic fields, and below a magnetization dependant critical temperature $T_{c1}$, a one component BEC (B phase) is obtained. Below a critical magnetic field $B_c$ \cite{noteBc}, corresponding to the field below which spin dependent contact interactions favor a non ferromagnetic ground state \cite{PhaseCr,Pasquiou}, we observe a multi-component BEC (C phase). Figure \ref{PhaseDiag} maps these results onto the phase diagram of non interacting spinor bosons as a function of temperature and magnetization \cite{Machida}.

The existence of the spin degree of freedom modifies statistics and therefore the thermodynamic properties of the gas, as $(2S+1)$ states come into play. Theoretical studies of non interacting spinor bosons \cite{Machida,Cohen} have been performed in presence of a purely linear Zeeman effect \cite{NoteQ}. They assume that collisions are fast enough for thermal equilibrium to be reached, but that interactions are weak enough that they do not modify the phase diagram. These studies generalize results obtained for scalar bosons, by introducing a Bose occupation factor for the non-condensed particles in each Zeeman state, leading to the following dependence for the thermal populations:

\begin{equation}
N_{th}^{m_S}\propto\sum_{n_x,n_y,n_z}\frac{1}{e^{\beta (\sum_{i=x,y,z} n_i\hbar\omega_i + m_S g_J \mu_B B -\mu)}-1}
\label{Nth}
\end{equation}
with $\omega_i$ the trap frequencies, $1/\beta=k_BT$, and $\mu$ the chemical potential. $B$ is either the experimental magnetic field if the magnetization is considered free, or, alternatively, an effective magnetic field self-consistently adapted to fix the magnetization thus behaving as a magnetic chemical potential.

Compared to the scalar situation, the general trend is a reduction of the critical temperature for condensation due to the increased number of degrees of freedom. The critical temperature of a polarized sample, $T_{c0}$, is reduced by a factor $(2S+1)^{1/3}$ (in a 3 dimensional harmonic trap) when the magnetization $M$ reaches 0. Two different regimes have been studied. When $M$ is fixed \cite{Machida} a double phase transition occurs: below a first critical temperature $T_{c1}(M)$ a polarized BEC emerges (phase B) i.e. only the lowest energy Zeeman component undergoes condensation, while all other components condense below a second critical temperature $T_{c2}(M)$ (phase C). This double phase transition is represented by a vertical line in the phase diagram of Fig \ref{PhaseDiag}. For a free magnetization \cite{Cohen}, BEC of a non-interacting gas only occurs in the lowest energy single particle state ($m_S=-3$ for $^{52}$Cr), which corresponds to the B phase: the BEC is always ferromagnetic, at any finite magnetic field, and the C phase is always avoided. In this regime the system follows the dashed line of Fig \ref{PhaseDiag}, the function $M(T)$ being calculated for a given magnetic field (here 1 mG), which maps the results of \cite{Cohen} onto the phase diagram of \cite{Machida}.

There are open questions regarding how such theoretical frameworks can describe $^{52}$Cr BECs at low magnetic fields: the magnetization is free, but the BEC is not ferromagnetic below $B_c$, as we have observed recently \cite{Pasquiou}. In this paper we show that the theory without interactions correctly describes thermodynamic properties of $^{52}$Cr gases, provided magnetization is considered free above $B_c$, and fixed below.

To perform our experimental study, we use forced evaporation to prepare a sample of typically 10 to 30 thousand $^{52}$Cr atoms, at temperatures ranging from 1.5 $\mu$K to 50 nK. These samples are produced in a crossed optical dipole trap \cite{BEC Cr}. We repeat the procedure for different values of the final energy cutoff set by the evaporative ramp, thus varying $T$; then the trap is recompressed at final trap frequencies independent of $T$. We control the $B$ field at the 100 $\mu$G level, and the magnetization of the cloud, $M=\sum_{i=-3}^3 p_i\times i$ (where $p_i$ is the relative population of the $m_i$ Zeeman state) is measured by a Stern-Gerlach procedure \cite{Pasquiou}. By varying the value of $B$ and $T$, we can investigate, after thermal equilibrium is reached, properties of a spin 3 gas for different magnetizations and temperatures.

\begin{figure}[h]
\centering
\includegraphics[width=2.7in]{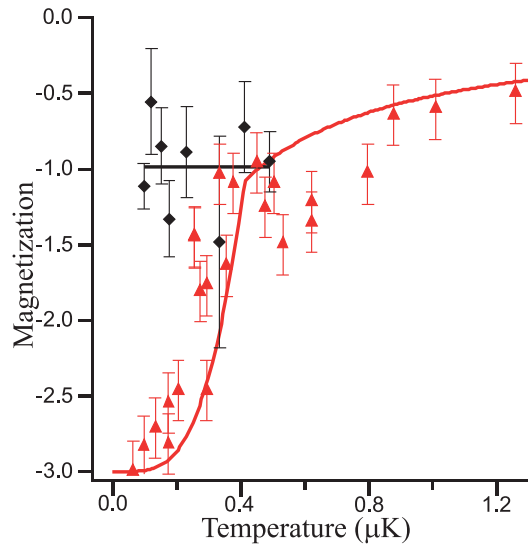}
\caption{\setlength{\baselineskip}{6pt} {\protect\scriptsize (Color online) Evolution of the magnetization with temperature. Triangles: above $B_c$ ($B=0.9$ mG), the magnetization $M$ slowly departs from zero as the temperature is lowered. When condensation is reached $M$ decreases faster, and reaches about $-3$ for the lowest temperatures; the data agree with a theory without interactions and free magnetization (solid line). Diamonds: below $B_c$ (here $B=B_{min}$), the magnetization remains almost constant and significantly different from $-3$, even at extremely low temperature (the horizontal solid line shows its average value).}} \label{Magnetization}
\end{figure}

For 6 different values of $B$, we have measured the steady state magnetization $M$ as a function of $T$, and the condensed fraction. We deduce six values of $T_{c1}$ (for which the condensed fraction departs from zero), which are reported in Fig \ref{PhaseDiag}. The observed decrease of $T_{c1}$ as $M$ increases from its minimal value $-3$ corresponds to a release of the spin degrees of freedom: the number of atoms in the most populated $m_S=-3$ state is lower than the total number of atoms, so that saturation is achieved at lower temperature. Fig \ref{PhaseDiag} shows good agreement between our results for $T_{c1}(M)$ and the theory of non interacting spinor bosons. Interestingly, the agreement between this simple model and the experimental results is good, while we explore values of $B$ both above and below $B_c$, corresponding to BEC of very different nature at $T=0$. This shows that $T_{c1}$ is relatively insensitive to interactions for our system, in both $B$ regimes; this is expected in the case of $S=1$ atoms \cite{Machida}.

Below $T_{c1}$, we find two entirely different behaviors of the system depending on the relative value of $B$ and $B_c$. This is illustrated in Fig \ref{Magnetization}, where we show experimental values of the gas magnetization as a function of temperature for two different magnetic fields. Above $B_c$, when temperature is decreased, the magnetization of the purely thermal (for $T>400$ nK) gas first progressively decreases from about zero (all the spin components being almost equally populated), to negative values, as the negative spin components become more populated. Below a certain temperature ($\approx400$ nK), we observe a kink in magnetization as well as a bimodal spin population distribution (see the experimental histogram characteristic of the B phase in Fig \ref{PhaseDiag}). At this temperature, a BEC is reached, as independently shown by the appearance of a bimodal momentum distribution measured after a time of flight. We interpret the kink in magnetization by the fact that only atoms in the $m_S=-3$ component condense, so that magnetization decreases faster with temperature. $M=-3$ is reached for the lowest temperatures, when almost all the atoms are Bose condensed. Fig \ref{Magnetization} shows the results of the theory without interactions and with free magnetization, and the good agreement with our experimental data. We hence pinpoint a spontaneous polarization of the cloud as condensation is reached, due to its ferromagnetic nature. Our data only reveal the equilibrium properties and it would be most interesting to study the spontaneous polarization of the condensed part due to DDIs if the temperature is lowered below $T_{c1}$ fast compared to magnetization dynamics.

\begin{figure}[h]
\centering
\includegraphics[width=3in]{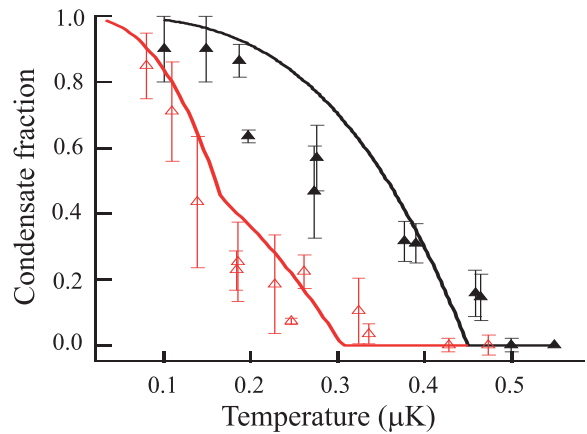}
\caption{\setlength{\baselineskip}{6pt} {\protect\scriptsize (Color online) Evolution of the condensed fraction $\eta$ with temperature. Filled triangles: for a polarized sample ($B=20$ mG), the usual $1-(T/T_{c0})^{3}$ (corresponding to the right solid line) is obtained. Open triangles: at $B_{min}$ (below $B_c$), the condensed fraction gets smaller at every temperature, as the spin degree of freedom is unfrozen. The left solid line corresponds to a theory with no interactions but fixed magnetization.}} \label{CondensedFraction}
\end{figure}

Below $B_c$ on the other hand, the results strongly depart from predictions of thermodynamics of non interacting bosons with free magnetization. We show in Fig \ref{Magnetization} results for the lowest magnetic field we can achieve ($B=B_{min}\leq 0.1$ mG). The magnetization does not reach $-3$, even for the lowest temperatures, whereas the condensate fraction reaches almost 1 in this situation (see Fig \ref{CondensedFraction}). Magnetization remains in fact almost constant, allowing to enter a C-like phase at low temperatures. The population distribution at the lowest temperature is represented by the lower histogram in Fig \ref{PhaseDiag}. For these data, we have checked that the thermal populations are negligible. It is worth noting that even though we enter a spinor phase, the spin distribution that we obtain is not the one predicted for the absolute ground state of chromium \cite{PhaseCr}. This may indicate that thermal equilibrium is not reached in this specific case, and that the BEC is in a metastable spin state.

A special emphasis is made for the temperature dependance of the condensed fraction $\eta$ below $B_c$ in Fig \ref{CondensedFraction}. The filled triangles correspond to data taken at large magnetic field ($B=20$ mG), while the open triangles correspond to $B=B_{min}<B_c$. At large $B$ field, the cloud remains polarized. In this case we measure a critical temperature $T_{c0}$, close to the value predicted for a non-interacting gas, as expected for weak interactions \cite{StringariRMP,Gerbier,Zoran}. For $B<B_c$ we observe a dramatic change: $\eta$ is significantly smaller, at any temperature below the critical temperature $T_{c0}$, than at high field. Large values of $\eta$ are only recovered for the lowest temperatures.

To explain these low $B$ fields results, we compare them with a theory with no interactions, but considering a fixed magnetization, as is empirically almost the case below $B_c$ (see Fig \ref{Magnetization}). Corresponding results are shown by the solid line in Fig \ref{CondensedFraction}, illustrating the double phase transition discussed in \cite{Machida}. In this experimental configuration, we measure $T_{c1}\approx300$ nK. Following the theory in \cite{Machida}, all Zeeman components should condense below $T_{c2}\approx150$ nK, and although we do not measure $T_{c2}$ we do find a multi-component BEC at the lowest temperatures \cite{Pasquiou}. The agreement between experimental data and the predictions of the model in Fig \ref{CondensedFraction} is satisfactory, hence showing a hint for the double phase transition. Thermodynamics of non interacting bosons with (almost) fixed magnetization hence seems to correctly describe our system. We however emphasize that a spinor C phase is only achievable below $B_c$ because then the ground state is not ferromagnetic, due to spin dependent contact interactions  \cite{Pasquiou}.

\begin{figure}[h]
\centering
\includegraphics[width=3in]{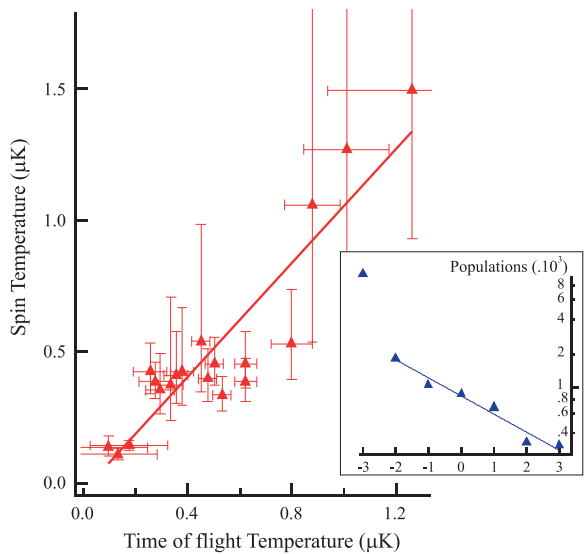}
\caption{\setlength{\baselineskip}{6pt} {\protect\scriptsize Comparison of the cloud temperatures deduced from a time of flight analysis ($T_{TOF}$) with those of a spin components analysis ($T_{spin}$), for $B=0.9$ mG. Inset: spin components populations. The magnetic field is set above $B_c$ ($B=0.5$ mG), so that only the thermal fraction gets depolarized. The temperature is below $T_{c1}$ ($T=160$ nK), hence the large $m_S=-3$ component. The other $m_S$ components almost obey a Boltzmann statistics \cite{noteBB}, as shown by the LogPlot, from which $T_{spin}$ is deduced.}} \label{Temperatures}
\end{figure}

Finally, we take advantage of results obtained for $B$ fields just above $B_c$ (i.e. in the B phase) in order to investigate a spin thermometry, efficient down to temperatures much smaller than $T_{c0}$. As explained above, for $B>B_c$ the BEC is only in $m_S=-3$, while the other $m_S$ components, purely thermal, obey a Bose statistics. A small value of $B$ allows thermal population in all $m_S$ components. We show in the inset of Fig \ref{Temperatures} the different populations measured for $B=0.5$ mG, at a temperature of 160 nK. The Bose statistics favors the negative values of $m_S$, while the large population in $-3$ emphasizes the ferromagnetic nature of the BEC. We derive the spin temperature, $T_{spin}$, by fitting the $m_S$ populations (for $m_S>-3$) to an exponential \cite{noteBB} .

In Fig \ref{Temperatures} we compare $T_{spin}$ to the temperature $T_{TOF}$ measured by a standard bimodal fit after a time of flight of 5 ms (without Stern-Gerlach separation). We observe a good general agreement between the two temperatures, indicating that spin and spatial degrees of freedom are at equilibrium. The error bars in $T_{spin}$ are a combination of the effect of the uncertainties on $B$ (dominant at low $T$), and of the statistical errors on the fit of the thermal populations \cite{notefit}. At $T$ way below $T_{c0}$, it is difficult to extract accurate temperatures from a bimodal fit of the velocity distribution  \cite{Davis}: the thermal cloud barely separates from the condensed part for $k_{B}T\leq\mu/3$, hence the increase in the error bars of $T_{TOF}$ at the lowest temperatures. At low $T$, $T_{spin}$ provides a more accurate measurement than $T_{TOF}$, showing the interest of this method.

This spin thermometry, relevant for dipolar gases, compares well with other techniques, such as noise thermometry with BECs coupled in a double well \cite{Oberthaler}. It bears similarities with spin thermometry using a magnetic field gradient \cite{KetterleSG}, but avoids the use of a gradient. As the minimal measurable temperature scales with $B_c$, this thermometry will therefore be most interesting for either ferromagnetic atoms, or atoms with small $B_c$; nevertheless, interactions between thermal atoms and the BEC could introduce systematics effects. Besides, one may take advantage of the selective depolarization of the thermal gas in the B phase to develop a cooling method, based on a configuration where only the $m_S=-3$ component is trapped: this would provide a selective loss process for the thermal cloud, which should result in cooling.

In conclusion, by working at low temperature and low magnetic field we have measured unique thermodynamic features of a gas of atoms with non zero spin, and free magnetization. We have evidenced strong differences in the behavior of the gas above or below $B_c$, the critical field separating ferromagnetic and unpolarized phases. We have mapped our results onto the phase diagram of a non interacting S=3 Bose gas. Below $B_c$ a metastable spinor C-like phase is observed. Above $B_c$, the BEC is always polarized, which we use to introduce a thermometry accurate even below the critical temperature.

\vspace{0.5cm}

Acknowledgements: LPL is Unit\'e Mixte (UMR 7538) of CNRS and
of Universit\'e Paris Nord. We acknowledge financial support from Conseil R\'{e}%
gional d'Ile-de-France, Minist\`{e}re de
l'Enseignement Sup\'{e}rieur et de la Recherche and
IFRAF. We thank Paolo Pedri for fruitful discussions.


\begin{thebibliography}{99}
\bibitem{noteBc} Although $B_c$ scales with the density of the condensed fraction, and is hence temperature dependent, $B_c$ will refer in the article to the value at zero temperature ($\approx 0.25$ mG), for sake of simplicity.
\bibitem{noteBB} Although the atoms follow the Bose statistics of (\ref{Nth}), in practice we use a Boltzmann fit, which induces negligeable systematic errors.
\bibitem{notefit} Here at large $T$, the statistical fit errors dominate (because populations in the different $m_S$ states are similar), but they can be reduced by working at larger $B$.
\end{thebibliography}
\end{document}